\documentstyle{article}
\newcommand{\vect}[1]{\mbox{\boldmath$#1$}}
\begin{document}
\begin{center}
{\LARGE \bf Signature and Angular Momentum in 3d-Cranked HFB states}\\
\vspace{5mm}
{\bf Makito Oi$^{\rm a}$\footnote{e-mail address: mon@nt2.c.u-tokyo.ac.jp}
, Naoki Onishi$^{\rm ac}$\footnote{e-mail address: onishi@onishi2.c.u-tokyo.ac.jp}, Naoki Tajima$^{\rm a}$\footnote{e-mail address: tajima@nt1.c.u-tokyo.ac.jp},
 Takatoshi Horibata$^{\rm bc}$}\footnote{e-mail address: horibata@aomori-u.ac.jp}\\
{\it $^{\rm a}$Institute of Physics, Graduate School of Arts and Sciences,
University of Tokyo, \\ Komaba, Meguro-ku, Tokyo 153, Japan}\\
{\it $^{\rm b}$Department of Information System Engineering,
Aomori University, \\ Kobata, Aomori 030, Japan}\\
{\it $^{\rm c}$Cyclotron Laboratory, Institute of Physical and Chemical
Research (RIKEN), \\
Hirosawa 2-1, Wako-city, Saitama 351-01, Japan}\\
\end{center}
\baselineskip 6mm
\begin{abstract}
In terms of the exact angular momentum projection,
properties of the three dimensional cranked HFB (3d-CHFB) states are 
analyzed quantitatively
in the context of the relation between the signature of an intrinsic 
symmetry and the parity of angular momentum, $(-1)^I$.
We found that the tilted states have favorable features to describe states
involved with high-$K$ quantum number and/or odd total angular momentum $I$.
This implies that 3d-CHFB can describe properly the backbending phenomena
like a ``t-band and g-band'' crossing, which is  suggested in $N$=106
isotones. 
\end{abstract}

In microscopic descriptions of rotational motions of atomic nuclei,
the cranked Hartree-Fock-Bogoliubov(CHFB) method has been a useful and
practical approach. It has been successful in analyzing
the regular rotational spectra [\ref{RS80}].

After the discovery of the backbending phenomena, the validity of the 
method for the application to this angular momentum region came
into questions [\ref{Ha76},\ref{GSF78}].
It is because the standard cranking model is based on
a semiclassical picture of a deformed wave packet uniformly rotating 
about an axis.
In such a state, the total angular momentum is highly mixed.
This mixing may not be a problem in regular rotational bands, 
in which the intrinsic state changes gradually with increase of 
angular momentum.
On the other hand, the backbending phenomenon involves
an abrupt change of structure
taking place in a rather short interval of angular momentum
due to a crossing of the ground-state band (g-band) with 
the rotational aligned band (s-band)
[\ref{SS72}].
Therefore, the cranking model may not be so adequate in the backbending 
region as the regular rotational bands.
Self-consistent treatments are expected to mend partially this shortcoming.

Hara, Hayashi and Ring [\ref{HHR82}] applied  the method of angular momentum
projection to the CHFB states to examine a quality of the states 
around the backbending region. 
Their studies are based on the cranking model for rotation 
about a principal axis of quadrupole deformation 
(principal axis rotation; PAR).
As we will show soon, the PAR-CHFB produces states having mainly even-$I$
and low $K$-quantum number components.
This implies that PAR-CHFB is a method analyzing rotational bands
associated with even-$I$.

With the intention of exploring yrare states
which involve high-$K$ and/or odd-$I$ states,
a three-dimensional self-consistent cranking model
was proposed [\ref{KO81}],
and a schematic calculation along the line was carried out
recently on the basis of HFB for $^{182}$Os [\ref{HO94}].
The calculation gives rise to interesting results such as 
a ``tilted axis rotating (TAR)'' state as a yrast state. 
In this paper, we study the  nature of 3d-CHFB states, which include both
PAR and TAR states, by means of the angular momentum projection. We discuss
whether the states are adequate for description of a new type of 
back bending phenomena caused by ``tilted rotation'' [\ref{WD93},\ref{PW97}].

Walker et al. [\ref{WD93},\ref{W95}] claimed from an experimental point of view
( in $^{180}$W and $^{182}$Os; isotones of N=106 ) 
that there is a high-$K$ band ($K^{\pi}=8^{+}$) interacting with the g-band 
and the s-band one after another at $I \sim 16$ 
in a quite short interval of angular momentum.
The yrast becomes the s-band at high spin 
and the high-$K$ band comes closer to the yrast.
It is possible that the high-$K$ band becomes the yrast before the s-band
comes down.
They speculate that the high-$K$ band can be a ``t-band''
because the Fermi energy of such nuclei is located in the middle of
high-$j$ (i$_{13/2}$) shell and the situation may result in ``Fermi alignment''
to produce TAR [\ref{Fr81}]. 
Such a new type of backbending caused by the ``g-t band crossing'' may
exhibit a characteristic pattern of the signature splitting 
and, possibly, a signature inversion.

We define in this paper the signature as the symmetry of the 
$\pi$-rotation about the 1-axis, which is one of principal axes of the
quadrupole moments. For PAR, the rotating axis coincides with the 1-axis;
the conventional definition of the signature [\ref{RS80}]. 
For TAR, the rotating axis is tilted by angle $\theta$ towards the 3-axis.

In our previous work [\ref{HOO95}], we estimated the signature splitting 
in terms of the generator coordinate method, in which
$\theta$ is chosen as a generator coordinate.
We defined the signature as the mirror symmetry with respect to the
equatorial plane.
In that  calculation employing one-dimensional generator coordinate,
our definition of the signature is equivalent to the present definition.
We assumed that the symmetric(antisymmetric) states are assigned 
to the even-$I$(odd-$I$) states.
This assignment is not obviously accepted and should be examined.
In this paper, it is our aim to evaluate quantitatively relationship
of signature and the parity of $I$ [\ref{T94}].

We construct the intrinsic state  $\left| \phi \right>$ based on
a variational calculation
with a generalized BCS wave function that is a vacuum of
the corresponding quasiparticles;
$ a_i \left| \phi \right> = 0. $
The creation and annihilation operators of the quasiparticles
$(a^{\dag}_i, a_i)$ are related to those of the nucleons
$(c_m ^{\dag}, c_m)$ via the generalized Bogoliubov transformation,
\begin{equation}
\left( \begin{array}{c} a_i \\ a^{\dag}_i \end{array} \right)= 
{\cal W^{\dag}}
\left( \begin{array}{c} c_m \\ c^{\dag}_m \end{array}\right), 
\quad {\rm with}~~~~~{\cal W}=\left( \begin{array}{cc} U_{mi}& V_{mi}^*\\ 
V_{mi} & U_{mi}^* \end{array} \right),
\end{equation}
where ${\cal W}$ is a unitary matrix.
Then, we solve the variational equation with constraints,
\begin{equation}
\delta \left< \phi \left|\left[ \hat{H} - 
{\textstyle \sum_{k=1}^{3}}\left( \mu_k\hat{J}_k 
 +\xi_k \hat{B}_k\right) - 
{\textstyle \sum_{\tau=\pi}^{\nu} }\lambda_{\tau}\hat{N}_{\tau} 
 \right]\right| \phi \right> = 0,
\end{equation}
where $\hat{J}_k$ and $\hat{B}_k$ is the $k$th component of total angular
momentum and the ``boost'' operators [\ref{KO81}], respectively, and 
$\hat{N}_{\tau}$ is a nucleon number operator.
The boost operators are expressed by mass quadrupole tensors $\hat{Q}_{ij}$
as,
\begin{equation}
\hat{B}_k = \frac{1}{2}(\hat{Q}_{ij}+\hat{Q}_{ji})~~~
 \quad (ijk; {\rm cyclic}),
\end{equation}
and the constraint reads,
$
\left< \phi \left| \hat{B}_k \right| \phi \right>=0,
$
acting to fix the principal axes of the quadrupole deformation.
In practice, we solve this equation by the method of steepest descent.
More detailed procedures to generate wave functions are presented
in Ref.[\ref{HO96}].

Next, we calculate angular momentum projection matrices.
\begin{equation}
\label{intg}
n^I_{KK'}(\theta,\theta ') = \left< \phi (\theta) \left| \hat{P}_{KK^{\prime}}^{I}\right|\phi (\theta ')\right>
~~~~{\rm with}~~~\hat{P}_{KK^{\prime}}^{I} =
\frac{2I+1}{8\pi^{2}}\int d\Omega D^{*I}_{KK'}
  (\Omega) \hat{R}(\Omega) .
\end{equation}
Here, $\hat{R}(\Omega)$ is a rotation operator through the Euler angles,
$\Omega \equiv (\alpha,\beta,\gamma)$, and
$ D_{KK^{\prime}}^I(\Omega)=
\langle IK \mid \hat{R}(\omega)\mid IK^{\prime}\rangle$ is the 
Wigner's function.
The states $\left| \phi (\theta)\right>$ and $\left| \phi (\theta ')\right>$ 
are CHFB solutions 
with tilting angles $\theta$ and $\theta '$, respectively.
Integration is written as,
\begin{equation}
  \label{intdef}
  \int d\Omega \equiv \int_0 ^{2\pi} d\alpha 
  \int_0 ^{\pi} \sin \beta d\beta
  \int_0 ^{2\pi} d\gamma.
\end{equation}

The overlap kernels, 
$\displaystyle \left< \phi (\theta) \left|\hat{R}(\Omega)\right|\phi (\theta')\right>$ ,
are evaluated by using the formulae [\ref{OH80}],
\begin{equation}
\label{oeq}
\left<\phi (\theta)|\hat{R}(\Omega)|\phi (\theta')\right> =\sqrt{{\rm det}\left|P(\Omega)\right|},
\end{equation}
where
\begin{equation}
\label{defp}
P(\Omega)=U(\theta)^{\dag}D^{\dag}(\Omega)U(\theta')
+V(\theta)^{\dag}D^{T}(\Omega)V(\theta').
\end{equation}
Calculation of the norm kernel has to be carried out with a special caution on
choosing the branches of the square root appearing on r.h.s. of eq.(\ref{oeq}).
In the present work, integration in eq.(\ref{intg}) has to be calculated 
straightforwardly due to the loss of symmetries, which are related to
 signature, reality of intrinsic states and conjugation of bra and ket. 
For details of these symmetries, see Ref.[\ref{HHR82}].

The intrinsic state and the projection operator
can be expanded in terms of  a complete orthonormal
set of angular momentum,
\begin{equation}
\left|\phi\right> = \sum_{IK\alpha} g_{K \alpha}^{I} \left|I K \alpha \right>
~~~~{\rm and}~~~\hat{P}^{I}_{KK^{\prime}} = 
 \sum_{ \alpha} \mid IK \alpha \rangle \langle  IK^{\prime} \alpha \mid~,
\end{equation}
where $\alpha$ indicates labels additional to $I$ and $K$.
The probability $w_{K}^{I}$ found  in $\left| IK\right>$
is written as,
\begin{equation}
w_{K}^{I}=\sum_{\alpha} \left| g_{K\alpha}^I \right|^2 = n^I_{KK},
\end{equation}
and therefore the probability to find states having a certain value of
$I$, is evaluated as,
\begin{equation}
W^I = \sum_{K=-I}^{I} w_{K}^{I} = Tr(n^I).
\end{equation} 

Then,
we carry out numerical calculations. 
Let us take a look at the nature of
the wave functions generated by 3d-CHFB.
Fig.1(a) and Figs.2(a,b) show the probability distributions of $I$ and $K$
in PAR states.
The PAR state is obtained under a constraint $\left< J_{x} \right>=13\hbar$,
which is in the band crossing region between the g-band and the 
s-band[\ref{HO96}].
One observes quite a regular distribution like a Gaussian in the present 
calculation.

We calculate  $\sum_{I} W^{I}$ up to $I=26$;
the sum to be 0.96 and 0.86 for $\left< \hat{J}_x \right>=$6 and 13, 
respectively.
We take only $36 \times 91 \times 36 =117936$ 
integral points 
$(\Delta \alpha =\Delta \gamma = 10^{\circ}, \Delta\beta=2^{\circ})$, 
so that we can not obtain values of $N^{I}_{KK}$
for very high $I$ and $\mid K \mid$ values.

In the PAR for $\langle \hat{J}_{x} \rangle = J$ and
$\langle \hat{J}_{z}\rangle =\langle \hat{J}_{y}\rangle = 0$,
there is a definite signature.
Therefore, one would expect that the state contains only even-$I$.
As seen in Fig.1(a), 
this holds quantitatively well. Even-$I$ components occupy $95\%$ for
$\langle\hat{J}_x\rangle=13\hbar$.
However, one can see that the odd
components are also included though their fractions are quite small.
This is attributed to $\gamma$-deformation associated with the 
self-consistent field;
The wave function for $\langle \vect{J} \rangle = 0$ contains
only even-$I$ states because $K=0$ and the $R_1$-symmetry.
The PAR states are generated by the operator 
$-\mu_{1}\hat{J}_{1}$, which
does not mix states of the different angular momenta.
However, our cranking procedure is self-consistent and induces
$\gamma$-deformation, which causes mixing of odd-$I$ states. 
From these facts it is clearly stated that the signature is not an exact
but effective quantum number to distinguish the states having even $I$
from odd $I$. 

In Figs.2(a,b) presenting the probability distribution $w_{K}^{I}$,
the intrinsic states consist mainly of the components with $K = 0$,
but there is a finite size of fluctuation around $K \sim 0$.
This fluctuation can be explained in terms of quantum fluctuation due to the
non-commutative nature among the angular momentum components
and the induced $\gamma$-deformation. 
Because the fluctuation is small,
the signature seems effective in PAR-CHFB states.
In this sense, it is reasonable to employ the PAR cranking model to
describe nuclear rotation in the yrast line.
However, it is obvious that the PAR intrinsic states are inappropriate 
for describing the backbending phenomena caused by ``g-t band crossing'' 
because of so small amount of odd $I$.

Fig.1(b) and Figs.2(c,d) illustrate the probability distribution of $I$
and $K$ in TAR states,
which are obtained under constraints on the total angular
momentum; $\left< J_1 \right>=13 \hbar \cos \theta, 
\left< J_2 \right>=0$ and $
\left< J_3 \right>=13 \hbar \sin \theta$.
Several characteristic features missing in the PAR states are found .
In Fig.1(b) presenting the probability distribution $W^{I}$,
one can see that more amounts of odd-$I$ components are contained 
in the TAR state than in the PAR. Conversely, the amount 
of the even-$I$ components is decreased to $83\%$ for $\theta=6^{\circ}$.
In Figs.2(c,d) presenting the probability distribution $w_{K}^{I}$
for $\theta=6^{\circ}$,
one can see that, although $K \sim 0$ components are still dominant,
$K \sim 6$ components are also sizable.
For a negative tilting angle, the second peak appears at $K \sim -6$.

These are explained by the fact that the TAR breaks the symmetry of
the signature of intrinsic state by tilting the rotating axis.
As a consequence, the TAR state contains high-$K$ components and 
odd-$I$ components as well as even-$I$ low-$K$ components.

 To restore the signature, 
we try a signature projection of the TAR states,
\begin{equation}
\left|\pm\right>= {\cal N} \left( \left|+\theta\right> 
                        \pm \left|-\theta\right>\right),
\end{equation}
where a normalization factor ${\cal N}$ is given by 
$1/\sqrt{2(1 \pm {\rm Re}\langle \theta \mid -\theta \rangle)}$
and $\left|\pm\right>$ express states
with ($\pm$)- signature.

After the signature projection, we achieve the angular momentum projection.
Fig.1(c) and Figs2.(e-h) show the probability distributions of $I$ and $K$
in signature projected TAR states.
For the projected state with (+)-signature, both profiles of $W^{I}$ and
$w_{K}^{I}$  resemble to those of the PAR state. Even-$I$ components 
occupy $89\%$.
One can understand this situation as that the tilting angle 
is small (6$^{\circ}$) and the PAR state has (+)-signature.
A different feature between them is that the projected state with 
(+)-signature contains more amount of high-$K$ components than the PAR.
The projected state with ($-$)-signature exhibits distinguished profiles 
from any of the other states. 
Even-$I$ and odd-$I$ components are mixed evenly ($50\%$ for the even-$I$) and 
$K\sim 5$ components appear.
Incidentally, in Figs.2(g,h), there are tiny peaks at $K=0$. These peaks are 
artifacts due to numerical errors.

When the overlap matrix $N_{K\,K^{\prime}}^{I}$ is diagonalized,
\begin{equation}
  \label{eigeneq}
   \sum_{K^{\prime}}N_{K\,K^{\prime}}^{I} g_{K^{\prime}}^{I \nu}
   = n_{\nu} ^I g_{K}^{I \nu},
\end{equation}
the eigenvalues have information on the degree of linear independence
of the wave functions, which is called multiplicity[\ref{KO77}].
The multiplicity is defined as,
\begin{equation}
    m(I)=\frac{(\sum_{\nu} n_{\nu} ^I)^{2}}{\sum_{\nu} (n_{\nu}^I)^2}
   =\frac{ (Tr[n^I])^2}{ Tr[(n^I)^{2}]}.
\end{equation}
In table 1, the probability $W^{I}$ , the multiplicity $m(I)$ and
the largest eigenvalues $n_1 ^I$ are shown for signature unprojected 
(designated by $(\theta)$),
(+)-signature and ($-$)-signature states.
In $W^I$, about four times larger odd-$I$ component is included 
in the ($-$)-signature state than the (+)-signature one while about twice 
larger even-$I$ components is included in the (+)-signature state than the 
($-$)-signature one. The multiplicity shows that every state,
except for odd-$I$ ($+,\theta$) states, consists 
mainly of a state with $n_1 ^I$
because values of the $m(I)$ are close to 1. 
Indeed, one can see that $W^I$ and $n^I _1$ are close in these cases.
\begin{center}
{\bf Table 1}
\end{center}

There are two conclusions from the present analysis. 
First, the signature is an effective quantum number associated with
an intrinsic symmetry to distinguish between even and odd $I$ states.
However, the resolution is not high enough to identify even-odd $I$
in practical discussion of signature splitting in ``g-t band crossing''.
The angular momentum projection is necessary in the study of yrare states.

Second, in the backbending region where the high-$K$ band crosses
with the g-band,
mixing of even-odd $I$ and high-$K$ quantum numbers are highly
expected.
Considering relative amounts of  even and odd total angular momenta,
TAR states are much more adequate for treating dynamics involving
odd-$I$ states than PAR states.

The numerical calculations are carried out by the Vector Parallel
Processor, Fujitsu VPP500/28 at RIKEN.
This work is financially supported in part by the Grant-in-Aid
for Scientific Research from the Ministry of Education,
Science, Sports and Culture of Japan (09640338)

\newpage
\begin{center}
  {\bf References}
\end{center}
\begin{enumerate}
  \setlength{\itemsep}{-0.1cm}
  \setlength{\parsep}{-0.1cm}
  \renewcommand{\labelenumi}{[\arabic{enumi}]}
\item\label{RS80}
  P. Ring and P. Schuck, ``{\it Nuclear Many-Body Problem}'' 
 (Springer-Verlag 1980) 
\item\label{Ha76}
 I. Hamamoto, Nucl. Phys. A271 (1976) 15
\item\label{GSF78} 
 F. Gr\"ummer et al., Nucl. Phys. A308 (1978) 77
\item\label{SS72}
F.S. Stephens and R.S. Simon, Nucl. Phys. A183 (1972) 257
\item\label{HHR82}
 K. Hara, A. Hayashi and P. Ring, Nucl. Phys. A358 (1982) 14
 \item\label{KO81} 
 A. Kerman and N. Onishi, Nucl. Phys. A361 (1981) 179,
N. Onishi, Nucl. Phys. A456 (1986) 279
\item\label{HO94} 
 T. Horibata and N. Onishi, Phys. Lett. B325 (1994) 283
\item\label{WD93}
 P.M. Walker et al., Phys. Lett. B309 (1993) 14
\item\label{PW97}
 C.J.Pearson et al., Phys. Rev. Lett. 79 (1997) 605 
\item\label{W95}
 P.M.Walker, private communications
\item\label{Fr81}
S. Frauendorf, Phys. Scr. 24 (1981) 349
\item\label{HOO95}
 T. Horibata, M. Oi and N. Onishi, Phys. Lett. B355 (1995) 433
\item\label{T94}
 N.Tajima, Nucl. Phys. A572 (1994) 365
\item\label{HO96}
 T. Horibata and N. Onishi, Nucl. Phys. A596 (1996) 251
\item\label{OH80}
 N. Onishi and T. Horibata, Prog. Theor. Phys 65 (1980) 1650
\item\label{KO77} 
 A. Kerman and N. Onishi, Nucl. Phys. A281 (1977) 373
\end{enumerate} 

\newpage
\begin{center}
  {\bf Figure Captions}
\end{center}
\begin{description}
\item[Fig.1] Probability distributions $W^I$.
 In each graph, even-$I$ and odd-$I$ are drawn separately. 
(a): PAR states obtained under the angular momentum constraints 
$\langle J_x \rangle = 6\hbar$(solid lines), $8\hbar$(dash-dot lines), $13\hbar$(dash-dot-dot lines) are compared.
(b): TAR states with $\theta = 0^{\circ}$ (solid lines), 
$2^{\circ}$ (dash-dot lines) and $6^{\circ}$ (dash-dot-dot lines)
$\do$ are compared. All of the states are obtained under 
$\langle J_x \rangle =13\hbar\cos\theta,
\langle J_y \rangle =0$ and $\langle J_z \rangle = 13\hbar\sin\theta$.
(c): Signature projected TAR states ($+/-$; solid/dash-dot-dot lines) and a TAR state ($\theta$; dotted line) 
are compared. Tilting angle is $6^{\circ}$ and the angular momentum constraints
are the same as in (b).
\item[Fig.2] Probability distributions $w_K ^I$.
(a): even-$I$ graph of a PAR state obtained under
$\langle J_x \rangle = 13 \hbar$.
(b): odd-$I$ graph of the same state as in (a).
(c): even-$I$ graph of a TAR state having $\theta=6^{\circ}$.
The angular momentum constraints are the same as in Fig.1(b).
(d): odd-$I$ graph of the same state as in (c).
(e): even-$I$ graph of a (+)-signature projected TAR state.
The angular momentum constraints and tilting angle are the same as those 
in Fig.1(c).
(f): odd-$I$ graph of the same state as in (e).
(g): even-$I$ graph of a ($-$)-signature projected TAR state.
The angular momentum constraints and tilting angle are the same as those 
in Fig.1(c).
(h): odd-$I$ graph of the same state as in (g).
\end{description}

\newpage
\begin{center}
{\bf Table caption}
\begin{description}
\item[Table 1] Probability $W^I$, multiplicity $m(I)$ and the largest eigenvalue of eq.(\ref{eigeneq}) $n_1 ^{\nu}$ are presented for each state;
$(\theta)$: a TAR state, $(\pm)$: $(\pm)$-signature projected states. 
Tilting angle is $6^{\circ}$, and the angular momentum constraint is; 
$\left|\langle \vect{J}\rangle\right| = 13\hbar$.  
\end{description}
\end{center}

\newpage
\begin{center}
{\bf Numerical values of probability, multiplicity and the largest eigenvalue}\\
\vspace{5mm}
\begin{tabular}{|l|ccc||l|ccc|}
\hline
  $I$( sign)   &  $W^{I}$  & $m(I)$ & $n_{1} ^I$ &
  $I$( sign)   &  $W^{I}$  & $m(I)$ & $n_{1} ^I$ \\
\hline
  11($\theta$) & 0.0153    & 1.93   & 0.011 &
  12($\theta$) & 0.0888    & 1.40   & 0.074 \\
  11($+$)      & 0.0108    & 2.64   & 0.006 &
  12($+$)      & 0.0960    & 1.22   & 0.087 \\
  11($-$)      & 0.0418    & 1.30   & 0.036 &
  12($-$)      & 0.0470    & 1.29   & 0.041 \\
  13($\theta$) & 0.0181    & 1.71   & 0.014 &
  14($\theta$) & 0.0889    & 1.46   & 0.072 \\
  13($+$)      & 0.0125    & 2.38   & 0.008 &
  14($+$)      & 0.0945    & 1.25   & 0.084 \\
  13($-$)      & 0.0511    & 1.24   & 0.046 &
  14($-$)      & 0.0555    & 1.26   & 0.049 \\
\hline
\end{tabular}
\end{center}
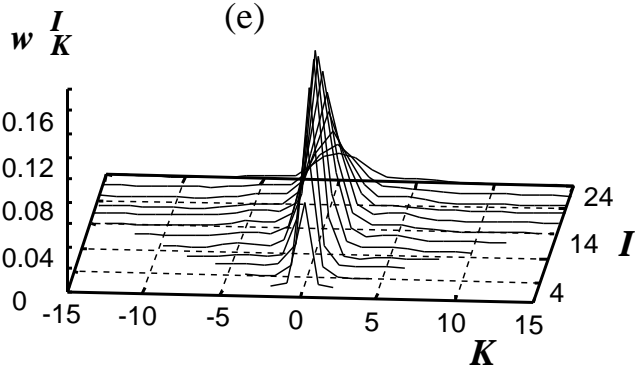
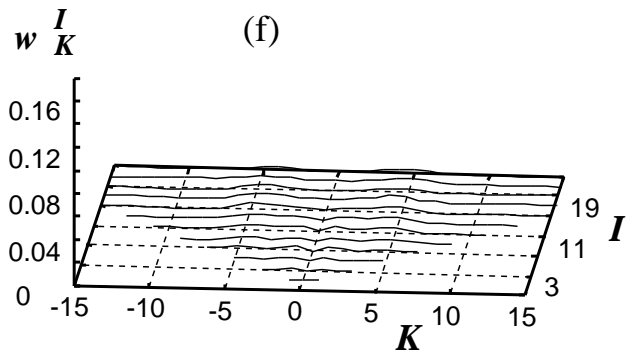
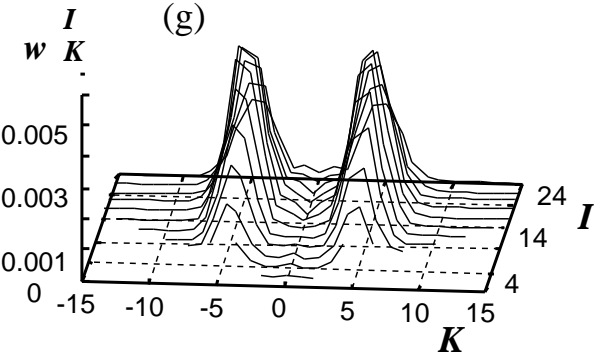
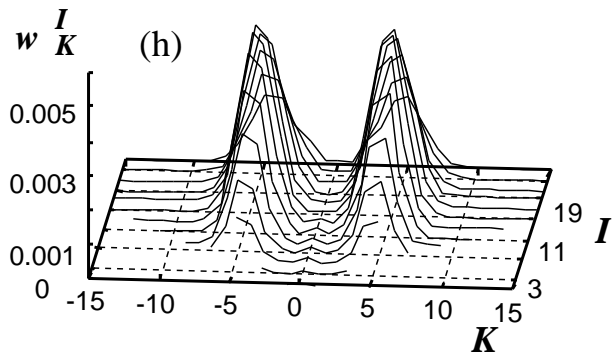
\end{document}